\documentclass[12pt,a4paper]{article}
\usepackage[width=.75\textwidth]{caption}
\usepackage{graphicx}
\usepackage{authblk}
\usepackage{amsmath}
\usepackage{amsfonts}
\usepackage{amssymb}
\usepackage{braket}
\usepackage{etoolbox}
\usepackage{orcidlink}
\usepackage[mathscr]{euscript}
\usepackage[top=2cm, bottom=2cm, left=2cm, right=2cm]{geometry}
\usepackage{fancyhdr}
\newcommand{\dv}[1]{\mathrm{d} #1 \text{ }}

\newcommand*\diff{\mathop{}\!\mathrm{d}}
\begin{document}

\title{Driving the Unruh Response}
\author[ ]{Kevin Player\footnote{\textit{kjplaye@gmail.com}} \,\orcidlink{0009-0000-7180-7985}}

\maketitle


\abstract{The Unruh effect, central to quantum field theory in curved spacetime, states that uniformly accelerated observers perceive the Minkowski vacuum as a thermal ensemble of Rindler excitations. Building on this foundation and drawing analogies from squeezing in quantum optics, we investigate how entangled, non-thermal excitations generated by bilocal driving sources contribute to the accelerated response. These paired excitations act as inertial microstates within the thermal Unruh ensemble, suggesting that portions of the effect can be interpreted as source-driven. To capture this, we employ modular automorphisms from algebraic QFT to track localization of modes and observers across nested Rindler wedges. We then construct compact wave-packet approximations using parabolic cylinder functions, providing a smooth interpolation between wedge-supported thermal modes and fully localized non-thermal excitations. This approach situates the Unruh response within a broader framework where standard thermality emerges alongside, and sometimes from, localized source-induced structure.}

\section{Introduction}

The Unruh effect \cite{unruh1976notes} reveals that uniformly accelerated observers describe the inertial Minkowski vacuum not as empty, but as a thermal bath of excitations. This apparent thermality reflects the deep entanglement structure of the vacuum: for inertial observers the Minkowski vacuum is pure and non-thermal, while for accelerated observers it appears as a mixed thermal state. Our work provides a complementary microstate-level understanding of this well-established theoretical framework.

Geometrically, uniform acceleration divides spacetime into left and right Rindler wedges, separated by an event horizon. The vacuum can be expressed as a product of entangled excitations across these wedges, so that tracing out one side yields a thermal description of the other. Local measurements in one wedge therefore have non-local implications for the other.

Recent work has refined the picture of Unruh thermality, highlighting how its form depends on detector locality, causal structure, and the balance between entanglement and information, and showing that, in each case, the correlations ultimately depart from strict thermality. Anastopoulos and Savvidou \cite{anastopoulos2012coherences} showed that while a single uniformly accelerated Unruh–DeWitt detector \cite{unruh1976notes,einstein1979general} exhibits thermal fluctuations, correlations between two spatially separated detectors with the same acceleration are non-thermal. This indicates that the Unruh effect is fundamentally local, with information encoded in correlations between distant observers. Foo, Onoe, and Zych \cite{foo2020unruh} studied detectors in quantum superpositions of classical trajectories. Although each trajectory individually yields a thermal response, the superposed detector does not thermalize, demonstrating the importance of causal relations rather than acceleration alone.

Svidzinsky, Scully, and Unruh \cite{Svidzinsky2024MinkowskiVEA} analyzed causal chains of harmonic oscillators spanning both wedges. They found that the entanglement of the Minkowski vacuum transfers directly to the entanglement of oscillators interacting with the field, reinforcing the causal underpinning of Unruh correlations. Han, Olson, and Dowling \cite{han2008generating} examined the role of measurement, showing that when an accelerating observer performs a projection-valued measurement (PVM), the field is driven out of its thermal character. For inertial observers the field is no longer vacuum, but contains real photons with nonzero energy–momentum, so the accelerated PVM acts as a process that injects energy and momentum through the accelerating agent itself.

In this work we develop a complementary viewpoint: rather then starting with acceleration and finding correlations, we investigate entangled emissions as a driver of the accelerated response. These driven excitations provide explicit examples of microstates consistent with the thermal Unruh ensemble, offering a minimal construction of how entangled sources can generate accelerated dynamics. We also map out generalizations by considering back-reaction, reversed emission and absorption, and related extensions.

In Section \ref{sec:prelim}, we review the Unruh effect, including the relevant mode expansions and Bogoliubov transformations. In Section \ref{sec:drive}, we construct a bilocal source, motivated by squeezed-state interactions in quantum optics, that injects correlated excitations into the field. Section \ref{sec:loc} explores partial localization by considering sub-regions of the Rindler wedge connected through space-like translations and reflections, transformations that correspond to modular automorphisms in the associated operator algebras. We then use parabolic cylinder functions to construct a smooth interpolation between eternal Rindler modes and fully localized excitations. Finally, in Section \ref{sec:future} we outline directions for future research, and in Section \ref{sec:conc} we interpret the implications of our construction.

\section{Preliminaries} \label{sec:prelim}

We draw notation and standard results from Frodden and Vald{\'{e}}s \cite{frodden2018unruh}. Let $\hbar$ = $c$ = 1. We work in 1+1 dimensional Minkowski spacetime with free, massless scalar fields, which capture the essential features of the Unruh effect without the overhead of higher dimensions or interactions. These simplifications preserve the relevant physics and clarify the constructions.

Consider the free scalar massless Lagrangian
\begin{equation}
\mathscr{L}_{\text{free}} = -\frac{1}{2} \eta^{\mu\nu}\partial_\mu \phi \partial_\nu \phi.
\end{equation}
We consider positive frequency modes with dispersion relation $\omega_k = |k| > 0$ as solutions to the resulting Klein-Gordon equation 

\begin{equation}
  \Box \phi = -\frac{\partial^2 \phi}{\partial t^2} + \frac{\partial^2 \phi}{\partial x^2} = 0,
 \label{massless-wave-eq}
\end{equation}
where $\Box = \eta^{\mu\nu} \partial_\mu \partial_\nu$. We expand $\phi$ in terms of ladder operators $a_k, a_k^\dagger$

\begin{equation}
  \phi(x,t) = \int \diff k \, a_k \varphi_k(x,t) + \text{h.c.}
\end{equation}
where

\begin{equation}
  \varphi_k(x,t) = \frac{1}{\sqrt{4\pi\omega_k}} e^{i(kx - \omega_k t)}.
\label{amode}
\end{equation}
are pure Minkowski positive frequency waves normalized with respect to the Klein-Gordon inner product over a Cauchy surface $\Sigma$ (usually $t = 0$)
\begin{equation}
  \left<f, g\right>_{KG} = i \int_\Sigma \diff x (f^* \partial_t g - \partial_t f^* g).
\end{equation}

\subsection{Rindler Coordinates} 

\begin{figure}[h]
\centering
\includegraphics[scale=0.2]{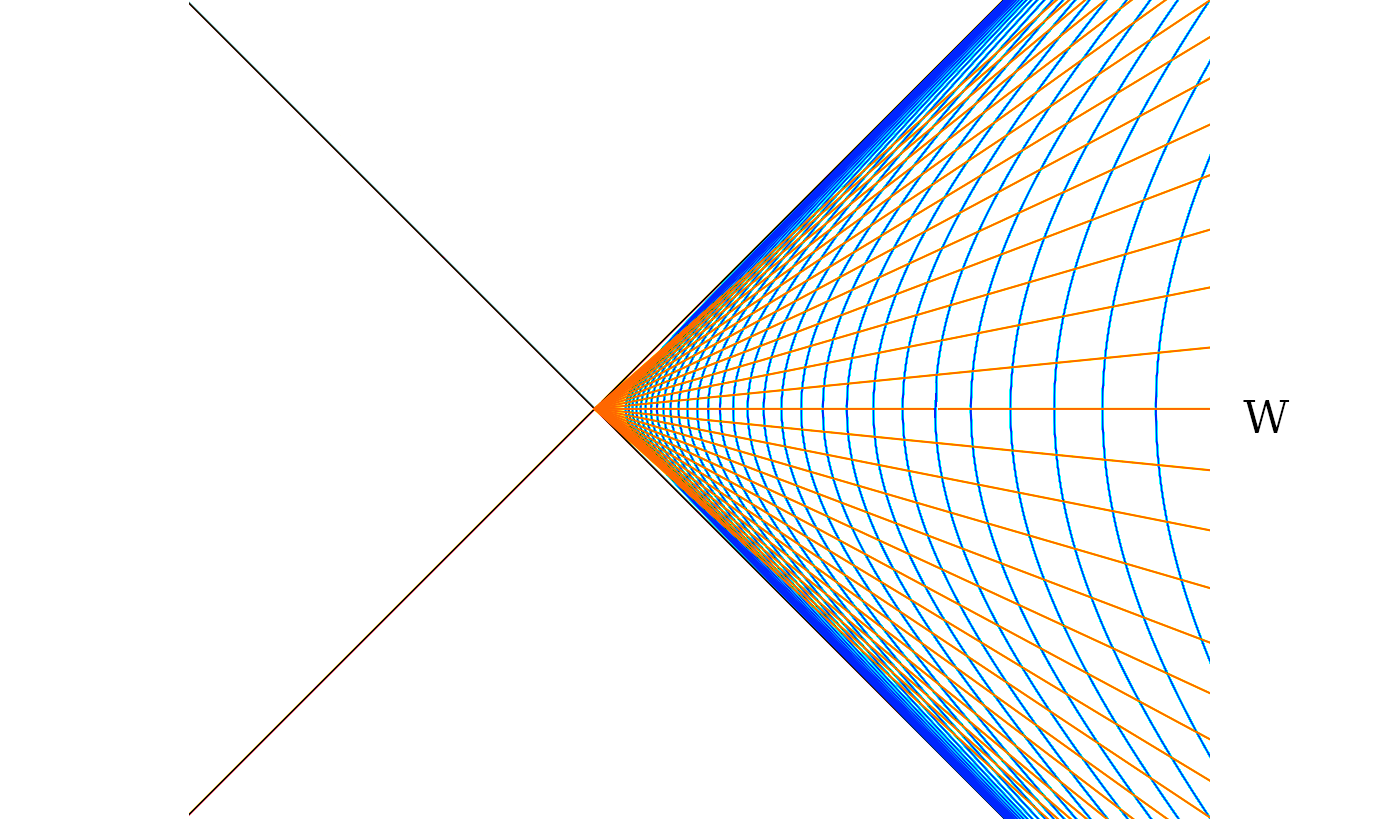}
\caption{Rindler wedge $W$ on the right, with Rindler coordinates.}
\label{rindlerw}
\end{figure}

To describe the physics from the point of view of a uniformly accelerating observer, we introduce Rindler coordinates \cite{frodden2018unruh,rindler1966kruskal} covering a right wedge 
\begin{equation}
  W = \{(x,t) : x>|t|\}
\end{equation}
with apex at the origin, pictured\footnote{All spacetime diagrams follow the convention of $t$ increasing upward and $x$ increasing to the right.} in Figure \ref{rindlerw}; with coordinates
\begin{equation}
  t = \frac{1}{a}e^{a\xi}\sinh{(a\eta)}
\label{sinh}
\end{equation}
\begin{equation}
x = \frac{1}{a}e^{a\xi}\cosh{(a\eta)}
\end{equation}
The constant acceleration parameter $a$ is introduced explicitly to make the dependence on the Unruh temperature, $T = \frac{a}{2\pi}$, manifest in subsequent expressions. The coordinates $(\eta, \xi)$ describe the proper time and spatial position of a uniformly accelerating observer, with constant $\xi$ corresponding to hyperbolic trajectories in Minkowski spacetime.

The massless Klein-Gordon equation in Rindler coordinates is
\begin{equation}
  \Box \phi = e^{-2a \xi}(-\partial_\eta^2 + \partial_\xi^2) \phi = 0
\end{equation}
The wave equation retains the same Minkowski structure up to the overall conformal factor $e^{-2a\xi}$. Since this factor does not affect the null structure of the equation, the mode solutions retain the same plane wave form but in the Rindler coordinates
\begin{equation}
 r_k(\eta,\xi) = \frac{1}{\sqrt{4 \pi \omega_k}} e^{-i(\omega_k \eta -k \xi)}
\end{equation}
for each wave number $k$ and positive frequency $\omega_k = |k| > 0$.  These Rindler modes are written in terms of $\eta$ and $\xi$ and are thus confined to the Rindler wedge $W$.  Since Rindler coordinates only cover $W$ (the right wedge), these modes are not defined globally in Minkowski space.

\subsection{Unruh Modes}
To review how a uniformly accelerated observer perceives the Minkowski vacuum as a thermal bath, we construct the Unruh modes\cite{unruh1976notes}, analytic continuations of Rindler modes that are positive-frequency\footnote{Positive frequency means that the modes contain no negative frequency Minkowski components.} solutions with respect to Minkowski time. From now on let $\omega_k = k > 0$.

We define constants $\alpha_k$ and $\beta_k$, directly tied to the thermal description, which satisfy $\alpha_k^2 - \beta_k^2 = 1$
\begin{equation}
  \begin{aligned}
    \alpha_k &= \frac{e^{\frac{\pi\omega_k}{2a}}}{\sqrt{2 \sinh \frac{\pi \omega_k}{a}}} = \sqrt{\frac{1}{1 - e^{-2\pi\omega_k / a}}} = \cosh \theta_k \\
    \beta_k &= \frac{e^{\frac{-\pi\omega_k}{2a}}}{\sqrt{2 \sinh \frac{\pi \omega_k}{a}}} = \sqrt{\frac{1}{e^{2\pi\omega_k / a} - 1}} = \sinh \theta_k \quad \text{(thermal form)} \\
  \end{aligned}
  \label{alpha_beta}
\end{equation}
where $\theta_k$ is defined so that $\tanh{\theta_k} = \frac{\beta_k}{\alpha_k} = e^{-\pi\omega_k / a}$. These show up throughout in mode normalizations\footnote{The normalizations come from computing the Klein-Gordon inner product on Minkowski space and comparing it to inner products on $W$ and $\widetilde{W}$.}, inner products, and resulting Bogoliubov transforms. $\beta_k$ also describes particle creation in terms of $|\beta_k|^2$ which has a thermal character, matching a Planck distribution at temperature $T = \frac{a}{2\pi}$.

\begin{figure}[h]
\centering
\includegraphics[scale=0.3]{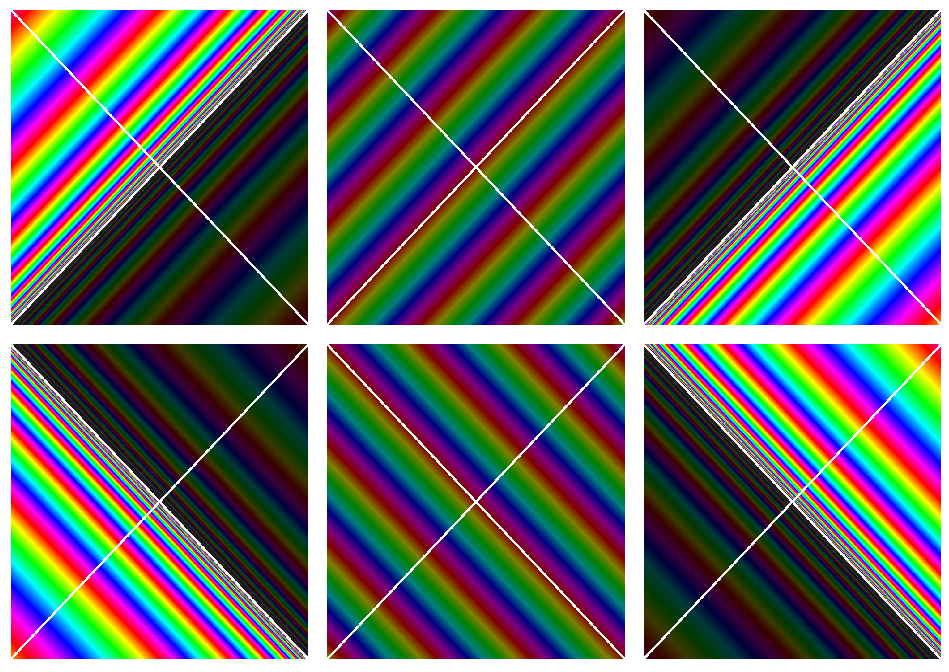}
\captionsetup{width=0.7\textwidth}
\caption{Spacetime diagrams of the $k>0$ mode functions $\left[\begin{array}{ccc} \mu^L_{-k} & \varphi_k & \mu^R_k \\ \mu^L_{k} & \varphi_{-k} & \mu^R_{-k} \end{array} \right]$ where $\mu$ and $\varphi$ are Unruh modes and Minkowski mode respectively. Color encodes the phase; brightness indicates magnitude.}
\label{unruh_rainbow}
\end{figure}

Let $\widetilde{W}$ be the left Rindler wedge\footnote{Coordinates on this wedge are $t = -\frac{1}{a}e^{a\delta} \sinh(a\gamma)$, $x = -\frac{1}{a}e^{a\delta} \cosh(a\gamma)$; and $l_{\pm k} = \frac{1}{\sqrt{4 \pi \omega_k}} e^{i\omega_k(\gamma \pm \delta)}$.}, $x < -|t|$ with Rindler modes $l_{\pm k}$.  We analytically continue\footnote{From the definitions and properties of $\sinh$ and $\cosh$, it follows that $a(\pm t + x) = e^{a(\pm \eta + \xi)}$ and then $r_{\pm k} = e^{\pm \frac{i \omega_k}{a}\log a(\mp t + x \pm i\epsilon)}$ see \cite{frodden2018unruh} for details.  Similar statements hold for the left wedge.} the Rindler modes $r_k$, $r_{-k}$, $l_k$ and $l_{-k}$ into the $(t,x)$ plane, these are the Unruh modes
\begin{equation}
  \begin{array}{ll}
    \mu^R_{\pm k} = \frac{\alpha_k }{\sqrt{4 \pi \omega_k}} (a(\mp t + x \pm  i \epsilon))^{\pm \frac{i \omega_k}{a}} & \hspace{20pt}
       {\mu^R_{\pm k}}_{|_W} \rightarrow \alpha_k r_{\pm k} \\
    \mu^L_{\pm k} =  \frac{\alpha_k}{\sqrt{4 \pi \omega_k}} (a(\mp t - x \pm  i \epsilon))^{\pm \frac{i \omega_k}{a}} & \hspace{20pt}
    {\mu^L_{\pm k}}_{|_{\widetilde{W}}} \rightarrow \alpha_k l_{\pm k} \\
  \end{array}
  \label{rindler_mode_def}
\end{equation}
An $i \epsilon$ prescription selects the branch of the logarithm that renders the modes analytic and bounded on the half plane $\Im(t)<0$. This makes the modes positive-frequency with respect to $t$.  Another way of writing the Unruh modes is
\begin{equation}
\begin{aligned}
  \mu^R_{\pm k} &= \alpha_k r_{\pm k} + \beta_k l^*_{\mp k} \\
  \mu^L_{\pm k} &= \alpha_k l_{\pm k} + \beta_k r^*_{\mp k} \\
\end{aligned}
\label{unruh_mode_def}
\end{equation}
where the right and left modes ($r_{\pm k}$ and $l_{\pm k}$) are understood to be zero outside of their respective wedges. See Figure \ref{unruh_rainbow} for an illustration of the various modes. The magnitude shown jumps across the branch cut and conjugates the phase. There are twice as many Unruh modes as Rindler modes, since each $r_k$ appears with two analytic extensions, $\mu^R_k$ and $\mu^{L*}_{-k}$, with a similar duplication for the left modes.

The Unruh modes form an alternative orthonormal basis of solutions to the Klein-Gordon equation, distinct from the plane waves $\varphi_{\pm k}$ see the original source, Unruh\cite{unruh1976notes}. The Unruh modes diagonalize (we will see in equation (\ref{diag})) the Minkowski vacuum in terms of Rindler particle states and thus provide the natural framework for describing the Unruh effect and the thermal response perceived by uniformly accelerated observers.

\subsection{Bogoliubov Transforms}
We generalize the wedge $W$ to a translated wedge $W_c$ with apex $(0,c)$
\begin{equation}
 W_c = \{(t,x) : x - c > |t|\} 
\end{equation}
and a reflected (left) wedge $\widetilde{W}_c$ with apex $(0,c)$
\begin{equation}
 \widetilde{W}_c = \{(t,x) : x - c < -|t|\}.
\end{equation}
Let the superscripts $(0)$, $(c)$, $(\widetilde{c})$, and $(M)$ represent the $W_0$, $W_c$, $\widetilde{W}_c$, and Minkowski frames of reference respectively.  Let $(A \rightarrow B)$ represent an open set inclusion\footnote{In the algebraic formulation of QFT, spacetime regions correspond to operator algebras. Here, we adopt a complementary (though formally contravariant) perspective, whereby shifts in the wedge induce Bogoliubov transformations between operator algebras.} $A \subseteq B$.

We directly compute $W_0 \rightarrow M$ Bogoliubov coefficients from equation (\ref{unruh_mode_def}) for a change of basis from $a^{(M)}_q$ to $c^R_q$ and $c^L_q$
\begin{equation}
  \phi = \int \dv{q} \mu_q^R c_q^R + \mu_q^L c_q^L + \text{h.c.}
  \label{c_ladder}
\end{equation}
We find a 4x4 block matrix with blocks of the form
\begin{equation}
  \left[ \begin{array}{l}
      a^{(0)}_k \vspace{10 pt}\\
    a^{\widetilde{(0)}\dagger}_{-k} \\
 \end{array} \right] = 
  \left[
\begin{array}{cc}
    \alpha_k &      \beta_k \\
    \beta_k        & \alpha_k \\
\end{array} \right]
\left[ \begin{array}{l}
    c^R_k \\
    c^{L\dagger}_{-k} \\
  \end{array} \right]
\label{diag}
\end{equation}
and three others, where the $a^{\widetilde{(0)}\dagger}_{-k}$ is a left wedge creation operator. We can summarize the transform in the right wedge $W$ as
\begin{equation}
  a_k^{(0)} = \alpha_k c_k^R + \beta_k c_{-k}^{L\dagger}
\label{a_in_c}
\end{equation}
and three other similar relations for $a_{-k}^{(0)}$, $a_{k}^{(0)\dagger}$, and $a_{-k}^{(0)\dagger}$.

We now compute the general non-diagonal Bogoliubov transformations.
\begin{equation}
  \begin{array}{rll}
  (c \rightarrow M) : & a^{(c)}_k &= \int \dv{q} \alpha^{(c \rightarrow M)}_{kq} a^{M}_q + \beta^{(c \rightarrow M)}_{kq} a^{(M)\dagger}_q \\
  (c \rightarrow 0) : &   a^{(c)}_k &= \int \dv{q} \alpha^{(c \rightarrow 0)}_{kq} a^{(0)}_q + \beta^{(c \rightarrow 0)}_{kq} a^{(0)\dagger}_q \\
  (\widetilde{c} \rightarrow 0) : &   a^{(\tilde{c})}_k &= \int \dv{q} \alpha^{(\widetilde{c} \rightarrow 0)}_{kq} a^{(0)}_q + \beta^{(\widetilde{c} \rightarrow 0)}_{kq} a^{(0)\dagger}_q \\
  \end{array}
\end{equation}
The $(c \rightarrow M)$ coefficients involve a gamma function, arising from the KG inner product as an integral of an exponential phase from $\varphi_k$ with a $(x-c)$ power from $r_k^{(c)}$ (the Mellin transform of $e^{ikx}$ \cite{bracewell1966fourier}):
\begin{equation}
  \begin{array}{ccl}
    \alpha^{(c \rightarrow M)}_{kq} &= \left<\varphi_q, r_k^{(c)} \right> &= \frac{1}{2 \pi a} \sqrt{\frac{\omega_k}{\omega_q}} \left(\frac{a}{q}\right)^{\frac{i\omega_k}{a}} e^{\frac{\pi \omega_k}{2a}} \Gamma\left(\frac{i\omega_k}{a}\right) \\
    \beta^{(c \rightarrow M)}_{kq} &= \left<\varphi_q^*, r_k^{(c)} \right> &= \frac{1}{2 \pi a} \sqrt{\frac{\omega_k}{\omega_q}} \left(\frac{a}{q}\right)^{\frac{i\omega_k}{a}} e^{\frac{-\pi \omega_k}{2a}} \Gamma\left(\frac{i\omega_k}{a}\right) \\
  \end{array}
  \label{bogoCM}
\end{equation}
Next we consider products of shifted powers to study $(c \rightarrow 0)$. We make use of a beta function for $(c \rightarrow 0)$ which occurs naturally in the KG dot product as an integral over a power of $x$ and of $x-c$, from $r_k^{(0)}$ and $r_k^{(c)}$ respectively.  We compute the Bogoliubov coefficients as
\begin{equation}
  \begin{aligned}
    \alpha^{(c \rightarrow 0)}_{kq} &= \left<r_q^{(0)}, r_k^{(c)} \right> = \frac{1}{2 \pi a}\sqrt{\frac{\omega_k}{\omega_q}} (ac)^{\frac{i(\omega_k - \omega_q)}{a}} B\left(\frac{i\omega_k}{a}, \frac{-i(\omega_k - \omega_q)}{a}\right) \\
    \beta^{(c \rightarrow 0)}_{kq} &= \left<r_q^{(0)*}, r_k^{(c)} \right> = \frac{1}{2 \pi a}\sqrt{\frac{\omega_k}{\omega_q}} (ac)^{\frac{i(\omega_k + \omega_q)}{a}} B\left(\frac{i\omega_k}{a}, \frac{-i(\omega_k + \omega_q)}{a}\right) \\
  \end{aligned}
  \label{bogoC0}
\end{equation}
The reflected diamond wedge version also yields a beta function, but with a different form
\begin{equation}
  \begin{aligned}
    \alpha^{(\widetilde{c} \rightarrow 0)}_{kq}     &= \left<r_q^{(0)*}, r_k^{(\widetilde{c})} \right> = \frac{1}{2 \pi a}\frac{\sqrt{\omega_k \omega_q}}{\omega_q - \omega_k} (ac)^{\frac{i(\omega_k - \omega_q)}{a}} B\left(\frac{i\omega_k}{a}, -\frac{i\omega_q}{a}\right) \\
    \beta^{(\widetilde{c} \rightarrow 0)}_{kq} &= \left<r_q^{(0)}, r_k^{(\widetilde{c})} \right> = \frac{1}{2 \pi a}\frac{\sqrt{\omega_k \omega_q}}{\omega_q + \omega_k} (ac)^{\frac{i(\omega_k + \omega_q)}{a}} B\left(\frac{i\omega_k}{a}, \frac{i\omega_q}{a}\right) \\
  \end{aligned}
  \label{bogoTC0}
\end{equation}

\subsection{Modular Automorphisms} \label{sec:mod}

Comparing absolute magnitudes for $M$ versus $W_c$ shows independence from $c$.
\begin{equation}
  \begin{array}{cc}
    \left|\alpha_{kq}^{(c_1 \rightarrow M)}\right|^2 = \left|\alpha_{kq}^{(c_2 \rightarrow M)}\right|^2 & \\
    \left|\beta_{kq}^{(c_1 \rightarrow M)}\right|^2 = \left|\beta_{kq}^{(c_2 \rightarrow M)}\right|^2 & \\
 \end{array}
\end{equation}
The $c$ independence is expected in this case since Unruh radiation is translation invariant. We next turn to $(c \rightarrow 0)$ and also find $c$ independence there 
\begin{equation}
  \begin{array}{c}
    \left|\alpha_{kq}^{(c_1 \rightarrow 0)}\right| = \left|\alpha_{kq}^{(c_2 \rightarrow 0)}\right| \vspace{4pt} \\
    \left|\beta_{kq}^{(c_1 \rightarrow 0)}\right| = \left|\beta_{kq}^{(c_2 \rightarrow 0)}\right| \\
  \end{array}
\end{equation}
This invariance is more surprising than in the Minkowski case, as it implies that the expected number of excitations for a mode $r_k^{(c_2)}$ when expressed in the vacuum of $W_{c_1}$,
\begin{equation}
  \int \dv{q} |\beta^{(c_2 \rightarrow c_1)}_{kq}|^2
\end{equation}
is invariant\footnote{Similar statements are true for reflected (diamond) wedges.} under changes in both $c_1$ and $c_2$.

More explicitly using the form of the $c$ term in equations (\ref{bogoC0}) and (\ref{bogoTC0}) we have a transform matrix of $\Lambda_c$ from $W_0$ to $W_c$ 
\begin{equation}
  \left[ \begin{array}{l}
    a^{(c)}_k \\
    a^{(c)}_{-k} \\
    \hline 
    a^{(c)\dagger}_k \\
    a^{(c)\dagger}_{-k} \\
 \end{array} \right] = \underbrace{
  \left[
\begin{array}{rr|rr}
    A_c        &       0   &  B_c            &  0 \\
    0        &      -A_c   &  0            & -B_c \\
    \hline 
    \overline{B_c}        &    0      &  \overline{A_c} & 0 \\
    0 &    -\overline{B_c}      &   0           & -\overline{A_c} \\
\end{array} \right]_{k,q} }_{\Lambda_c}
  \left[ \begin{array}{l}
    a^{(0)}_q \\
    a^{(0)}_{-q} \\
    \hline
    a^{(0)\dagger}_q \\
    a^{(0)\dagger}_{-q} \\
 \end{array} \right]
\end{equation}
where $A_c = \alpha_{kq}^{(c \rightarrow 0)} = P_c A_1 P_c^{-1}$  and $B_c = \beta_{kq}^{(c \rightarrow 0)} = P_c B_1 P_c$ for a diagonal phase factor matrix
\begin{equation}
  P_{c,rs} = \delta(r - s) c^{\frac{i\omega_r}{a}} = e^{\frac{i}{a} H \log c}
\end{equation}
where $H$ is the Rindler Hamiltonian associated with mode frequency $\omega_k$. We can write $\Lambda_c$ out compactly as
\begin{equation}
  \Lambda_c = Q_c \Lambda_1 Q_c^{-1}
\end{equation}
where
\begin{equation}
  Q_c = \left[\begin{array}{cccc}
        P_c, & 0 & 0 & 0 \\
        0 & P_c & 0 & 0 \\
        0 & 0 & P_c^{-1} & 0 \\
        0 & 0 & 0 & P_c^{-1} \\
    \end{array} \right] 
\end{equation}
The composition of Bogoliubov transforms, $\Lambda_{nc} = \Lambda_c^n$, yields
\begin{equation}
  \begin{array}{ll}    
    Q_{nc} \Lambda_1 Q_{nc}^{-1}  &= \Lambda_{nc} \\
         &= \left(Q_c \Lambda_{c} Q_c\right) \left( Q_c^{-1} \Lambda_{c} Q_c\right) \cdots \left(Q_c \Lambda_{c} Q_c\right) \\
  &= Q_c \Lambda_c^n Q_c^{-1} \\
  \end{array}
\end{equation}
so that
\begin{equation}
  \begin{array}{ll}
  \Lambda_c^n &= Q_c^{-1} Q_{nc} \Lambda_1 Q_{nc}^{-1} Q{c} \\
  &= Q_n \Lambda_1 Q_n^{-1}
  \end{array}
\end{equation}
and more generally we have a one parameter unitary group under the modular parameter $x = \log c$, with generator $\frac{1}{a} H$ given by
\begin{equation}
  \left\{\Lambda_1^x = Q_x \Lambda_1 Q_x^{-1} : x \in \mathbb{R} \right\}.
\end{equation}
Thus the Bogoliubov transformations between shifted wedges form a one-parameter group under translations of the apex, paralleling modular automorphism flow in algebraic QFT \cite{borchers2000revolutionizing}. In contrast to traditional treatments emphasizing Lorentz boosts within a fixed wedge, this formulation reveals modular structure via spatially translated wedges.

Consider a sequence
\begin{equation}
  W_{c_n} \subsetneq \cdots \subsetneq W_{c_i} \subsetneq \cdots \subsetneq W_{c_j} \subsetneq W_{c_2} \subsetneq W_{c_1}
  \label{chain}
\end{equation}
Each inclusion $W_{c_i} \subsetneq W_{c_j}$ yields the same particle production, with fixed squared Bogoliubov magnitude $|\beta_{kq}|^2$, so the expected number of particles remains constant across all nested wedge pairs, independent of the specific values of $c_i$ or $c_j$. We will briefly revisit this chain in Section \ref{sec:chain_sources}. This structure is closely related to the causal chains within the wedges discussed in \cite{Svidzinsky2024MinkowskiVEA}.

\section{Driving Sources} \label{sec:drive}

We now turn to a foundational question -- {\it what, physically, accelerates the observer?} In many treatments, including our own preliminaries in Section \ref{sec:prelim}, acceleration enters as a geometric input, a coordinate choice, with no reference to an underlying driving mechanism. Moreover, we have left unspecified both the observer’s precise location within the Rindler wedge and the spatial origin of the detected excitations. These omissions reflect an effective coarse-graining over the details of the observer and their interaction with the field, features that contribute to the apparent thermality observed in the Unruh effect.

\begin{figure}[h]
\centering
\includegraphics[scale=1.0]{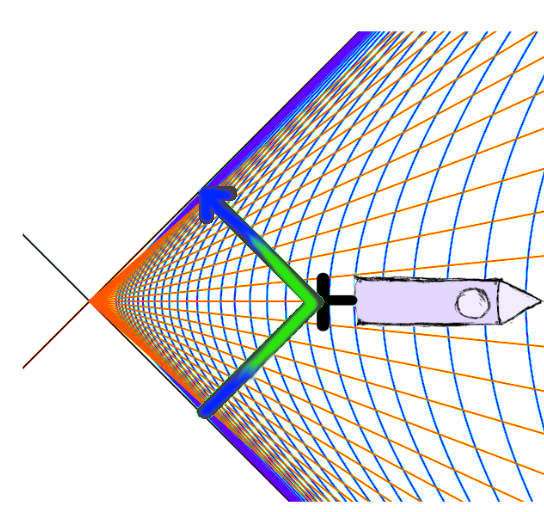}
\captionsetup{width=0.7\textwidth}
\caption{A Rindler mode's frequency is smeared out in Minkowski space, blueshifted near the horizon, and redshifted as $x$ goes to positive infinity. We illustrate a particle striking a mirror at the rear of a rocket, where its reflection emerges as a combination of emission and absorption processes in the Rindler frame.}
\label{emit_absorb}
\end{figure}

A natural physical interpretation is that a driving source must exist, both as the cause of the observer’s acceleration, and as a classical source coupled to the quantum field. The source results in active, localized interactions along the observer’s worldline, responsible for the observer’s motion.  This aligns with the idea that the observer is not in an isotropic background radiation, but is actually accelerated away from a thermal event horizon, the thrust being intimately correlated with the causal horizon.

Figure \ref{emit_absorb} illustrates the situation with a particle composed of Rindler modes on the right wedge. The modes $r_k$ are right-moving, originating from the past horizon and associated with emission; the $r_{-k}$ modes are left-moving, propagating toward the future horizon and are associated with absorption. These Rindler modes are constructed as superpositions of restricted Minkowski modes ${\varphi_q}_{|_W}$ effectively smeared across a range of frequencies.  This frequency mixing is evident in Figure \ref{unruh_rainbow}, where the modes blueshift infinitely near the horizons and redshift infinitely at spatial infinity, reflecting the geometry of the wedge.

\subsection{Construction}

\begin{figure}[h]
\centering
\includegraphics[scale=0.75]{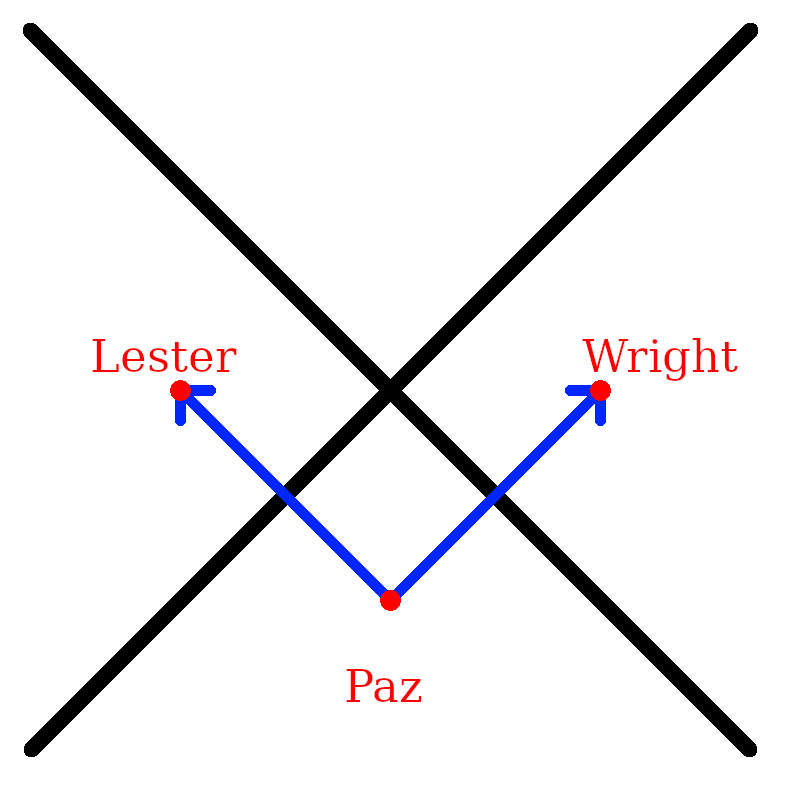}
\captionsetup{width=0.7\textwidth}
\caption{A source in the past, Paz(P), emits an entangled particle pair that accelerates Wright(R) (to the right) and Lester(L) (to the left).}
\label{paz}
\end{figure}

As a minimal illustrative construction we adopt an Alice-and-Bob style naming convention and use the names Wright (R), Lester (L), and Paz (P) to represent observers at $(t,x)$ = $(0,1)$, $(0,-1)$, and $(-1,0)$, respectively, see Figure \ref{paz}. These labels indicate Right, Left, and Past, serving as a mnemonic for their positions in spacetime. Paz serves as a common causal ancestor to Wright and Lester, and emits an entangled pair of particles\footnote{Strictly speaking, these are spin 0 massless scalar quanta.} with momentum $\pm k$ and (Rindler) frequency $\omega = \omega_k$ as part of a stream that, in our model, provides the correlated accelerations of Wright and Lester.  Note that we purposefully pick a single, sharply peaked $k$ and do not sample it from a thermal distribution.

Because the emitted particles carry perfectly balanced momenta, our basic model exhibits no back reaction -- this is a simplification -- in more general models, single-particle sources would encode back reaction through their entanglement structure, see Section \ref{sec:drive_phy} and Figure \ref{back_react}.

We next invert the usual logic: rather than coupling a detector to a pre-existing acceleration and deriving a thermal profile, we instead place an interaction in the past, where a source injects non-thermal entangled excitations into the field. The setup is prescriptive; by construction, the detector is guaranteed to respond to the momentum $k$ particle. In this picture, the observer’s accelerated response is driven by these source-induced excitations. This perspective does not contradict detector-based approaches but complements them, highlighting entangled emissions as a causal origin for part of the Unruh response. 

In this way, Paz's emission makes the source of the acceleration manifest through particle injection, so that Wright will necessarily detect a particle with peaked right-moving momentum $k$. Wright’s detection is entangled with Lester’s, and Lester can also observe a particle with left moving momentum $-k$. In this picture, Paz's emission acts as a bilocal source 
\begin{equation}
  \begin{aligned}
    J(x,y) &= \lambda f_L(x)f_R(y)\\
    \mathscr{L}_{\text{sourced}} & = \mathscr{L}_{\text{free}} + \frac{1}{2}\lambda \int f_L(x)f_R(y) \phi_+(x) \phi_+(y)  + h.c.\\
                        &= \mathscr{L}_{\text{free}} + \frac{1}{2}\lambda \left(q_{-k}^{L\dagger} q_{k}^{R\dagger} + q_{-k}^{L} q_{k}^{R} \right)
 \end{aligned}
 \label{bilocal}
\end{equation}
with coupling constant $\lambda$, normalized mode profiles $f_L$ and $f_R$ supported on the left and right wedges respectively, creation operators $q_{-k}^{L\dagger}$ and $q_{k}^{R\dagger}$, and $\phi_+$ as the positive frequency (annihilator) part of $\phi$. The last part of equation (\ref{bilocal}) is the squeezing interaction for frequency $\omega$, so that the bilocal source creates entangled particles in the Rindler basis within the wedges.

This source construction is a quadratic generalization of the usual linear source in QFT \cite{Schwinger_1966} \cite{ryder1996quantum} and is motivated by quadrature squeezed light generation in quantum optics \cite{gerry2023introductory}; it is a coupling of a classical function $J$ with the field as an interaction term smeared over $x$ and $y$. In this model $J$ squeezes the field creating correlations that manifest as entanglement at the level of a global quantum state. The product structure reflects this paired excitation.

We consider the case where $f_L$ and $f_R$ are absorptions, or made up of left moving and right moving Minkowski modes respectively; the emission case is similar, see Section \ref{sec:drive_phy}. In this picture, the bilocal source is not just a mathematical device but the physical mechanism by which acceleration is generated: the entangled pair created by $J(x,y)$ supplies the correlated excitations that are experienced as thrust by the Rindler observers.

Smearing with $\phi_+$ projects $f_L$ and $f_R$ onto their positive frequency parts which we call $g_L$ and $g_R$ respectively. In particular we can choose $f_L$ and $f_R$ to be Rindler modes $l_{-k}$ and $r_{k}$ in which case the projection is onto the corresponding Unruh modes $\mu_{-k}^L$ and $\mu_{k}^R$, so that $q_{-k}^{L\dagger}$ and $q_{k}^{R\dagger}$ are the corresponding Unruh operators $c_{-k}^{L\dagger}$ and $c_{k}^{R\dagger}$\footnote{See Section \ref{sec:prelim} equations (\ref{rindler_mode_def}), (\ref{unruh_mode_def}), and (\ref{c_ladder}).}.  

\begin{table}[ht]
\caption{Bilocal Source -- Modes, Operators, and States}
\centering
\begin{tabular}{c c c c c} 
\hline\hline
Type & Source & Projection & Operator & State (1st order)\\ [0.5ex] 
\hline 
General Left & $f_L$ & $g_L$ & $q_{-k}^{L\dagger}$ & $\ket{\psi}_{L}$\\
General Right & $f_R$ & $g_R$ & $q_{k}^{R\dagger}$  & $\ket{\psi}_{R}$\\
Rindler/Unruh Left & $l_{-k}$ & $\mu_{-k}^L$ & $c_{-k}^{L\dagger}$ & $\ket{1_\omega}_{L}$\\
Rindler/Unruh Right & $r_{k}$ & $\mu_{k}^R$ & $c_{k}^{R\dagger}$ & $\ket{1_\omega}_{R}$\\
\hline 
\end{tabular}
\label{table:modes}
\end{table}

We next consider the standard thermal vacuum description\footnote{The form of this vacuum equation can be proven by recursively applying the Bogoliubov relations from equation (\ref{a_in_c}).} of the Unruh effect
\begin{equation}
  \ket{0}_M = \prod_{\omega} \frac{1}{\cosh{\theta_k}} \sum_n \tanh^n{\theta_k} \ket{n_\omega}_{L} \ket{n_\omega}_{R}.
  \label{vacuum}
\end{equation}
A state that factorizes into correlated excitations in the left (L) and right (R) wedges. The Unruh modes associated with fixed frequency $\omega$ are specific linear combinations of these $L$ and $R$ Rindler modes, so that any source $J$ coupling to it will, in general, drive correlated excitations across both wedges. For example, the state prepared by Paz with Rindler modes corresponds to a selective excitation at a fixed $\omega$,
\begin{equation}
  \ket{\Psi_\omega}_M = \ket{1_\omega}_L \ket{1_\omega}_R + \text{O(higher terms)},
  \label{one_fock}
\end{equation}
a controlled realization of one term in the ensemble coming from the Rindler/Unruh modes. Our source projects the state onto the $\omega$ term through the act of observation, i.e. by a PVM projection in the standard quantum-measurement sense (see also \cite{han2008generating}), so that Wright’s guaranteed detection is modeled as the measurement-induced selection of a specific component of the thermal ensemble. The source $J(x,y)$ does not contradict the thermal interpretation, but rather describes a specific microstate consistent with the broader thermal ensemble.

\begin{figure}[h]
\centering
\includegraphics[scale=0.5]{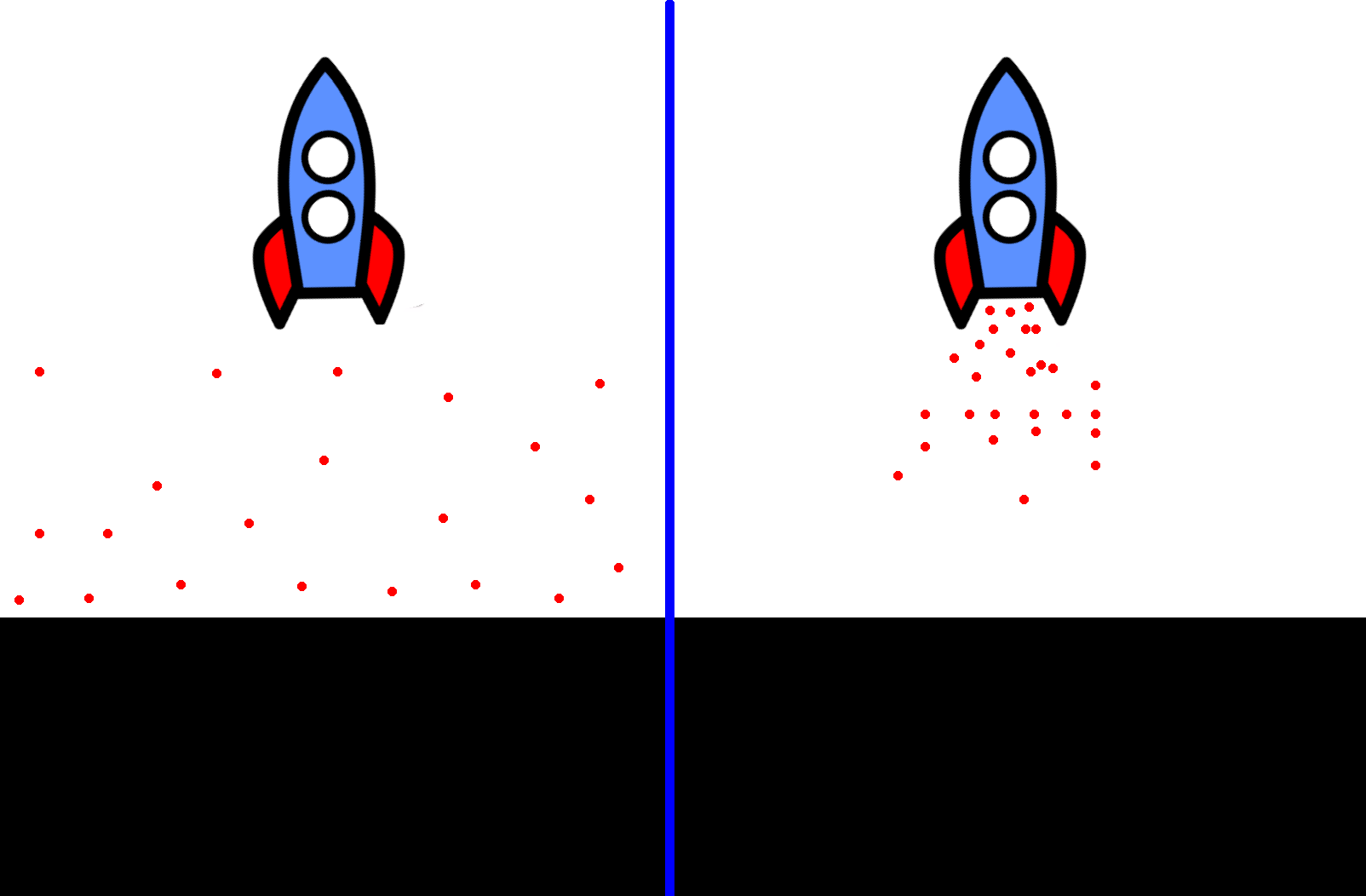}
\caption{Conceptual illustration contrasting thermal acceleration (left) with localized, source-driven acceleration (right). A portion of the rocket's acceleration arises from spatially localized driving sources rather than diffuse thermal effects.}
\label{rocket_inertial}
\end{figure}

An illustration of this setup is shown in Figure \ref{rocket_inertial}. From this perspective, the apparent thermality arises from intrinsic properties of the vacuum, an effective ignorance of the source's detailed structure and dynamics. In the sourced view, a portion of the Unruh effect is not a passive revelation of hidden particles in the vacuum, but a measurable consequence of thrust.

\subsection{Physical Details} \label{sec:drive_phy}

The construction of a right moving absorption function $f_R$ for Wright presents some physical challenges.

\begin{itemize}
\item {\bf Detector Modeling}. In standard treatments, thermal features are inferred from the response of an Unruh--DeWitt detector \cite{unruh1976notes,einstein1979general} coupled to the field. In the present framework, excitations are instead injected directly at the source. A systematic comparison of source-driven excitations with detector-based response functions could clarify how localized driving modifies, complements, or modulates thermal detector statistics, but for now it is beyond the scope of this paper.
  
\item {\bf Mode Projection}. We note that the $f_L$ and $f_R$ that match equation (\ref{one_fock}) are exactly the Rindler modes $l_{-k}$ and $r_{k}$ from Section \ref{sec:prelim}. The source $J$ is described in the Rindler wedges, and not immediately on the past wedge, so Paz will need to couple to Unruh modes -- the analytically continued Rindler modes.  We form $f_R$ out of a linear combination of a left Unruh mode $\mu^{L*}_{-k}$ and right Unruh mode $\mu^R_{k}$ to get $r_{k}$ supported on just the right wedge, see equation (\ref{unruh_mode_def}).  The projection in equation (\ref{bilocal}) encodes this relation by projecting the Rindler modes onto Unruh modes or in more generality by projecting $f_L$ and $f_R$ onto positive frequency $g_L$ and $g_R$.

\item{\bf Source over Finite Time}. We are integrating over the entire Minkowski space instead of over a short time period as normally desired for a source $J$.  This challenge is met by noticing that the projected positive frequency modes $g_L$ and $g_R$ can be restricted and scaled to a small time interval $|t + 1| < \delta$ around Paz's time $t=-1$ to give the same result, see Figure \ref{paz_freq} (right picture). This is because the integral over $x$ is independent of $t$. This is manifestly evident by the fact that the mode is translation invariant over time and space, e.g. $g_R(t,x) = g_R(-1,x-t-1)$, since it is a sum of right moving Minkowski modes. The coupling constant $\lambda$ in equation (\ref{bilocal}), along with $\delta$, can be chosen to match the vacuum equation (\ref{vacuum}) at a fixed $\omega$.

\begin{figure}[h]
\centering
\includegraphics[scale=0.5]{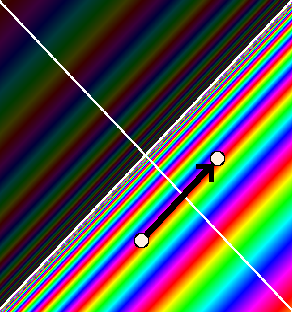}
\hspace{20 pt}
\includegraphics[scale=0.5]{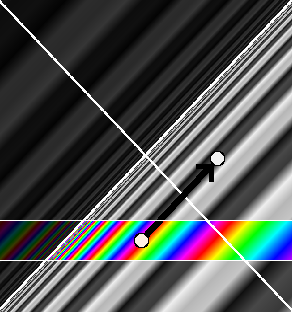}
\captionsetup{width=0.7\textwidth}
\caption{The points marked with white dots at (0,1) (Wright) and (-1,0) (Paz) respectively in the left picture lie on the same Unruh mode trajectory; although the mode is blueshifted infinitely  at the horizon, it settles down at the null line between Wright and Paz's locations. The right figure illustrates a finite time interval of width $\delta$ for the source $J$ to be active. Note how each time slice is a translation of the time = -1 slice.}
\label{paz_freq}
\end{figure}

\item {\bf Blueshifting near Horizon}. A third note is that the Unruh modes undergo an infinite blueshift at the horizon, but are more well behaved at Wright and Paz's locations\footnote{$f_R$ is actually zero at Paz and in fact on the entire past wedge, but the projection $g_R$ is an Unruh mode which is nonzero at Paz's location.}. We study the localization of functions on the right wedge in subsequent sections where we aim to attenuate or bound $f_R$ away from the high-frequency, near-horizon behavior typical of thermal modes, making the physical realization of the source $J$ more plausible as a wave packet.

\begin{figure}[h]
\centering
\includegraphics[scale=0.75]{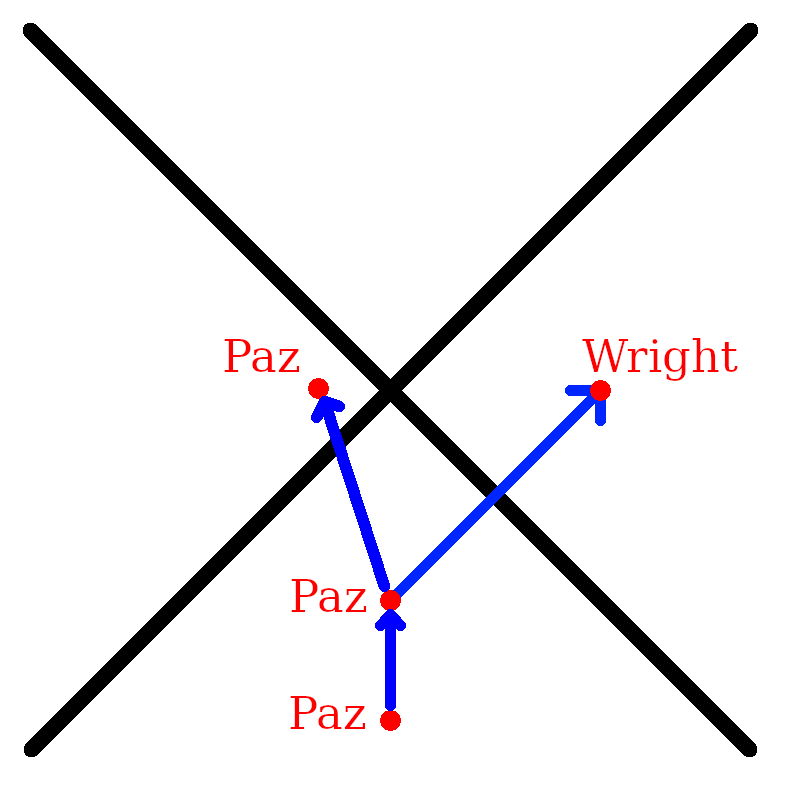}
\captionsetup{width=0.7\textwidth}
\caption{Paz emits a single particle and experiences a back reaction pushing her toward the left wedge. Through recoil she becomes entangled with the emitted particle, effectively taking on the role that Lester played in the symmetric construction.}
\label{back_react}
\end{figure}

\item {\bf Single Particle Models}. We mention that our running example represents just one of many possible mechanisms for exchanging energy and correlations between wedges; for example, in our symmetric two-mode setup, the momenta are balanced and the source does not experience recoil, so there is no back reaction. By contrast, if Paz emitted a single particle, as in Figure \ref{back_react}, the emitted mode would necessarily be entangled with Paz’s internal state and/or recoil momentum, producing back reaction as field–source entanglement. Tracing out the source would leave Wright with a mixed state, and the apparent thermality would then arise from this entanglement with the source’s degrees of freedom. Our choice of a balanced bilocal source thus isolates the thrust mechanism while postponing the additional complication of explicit source recoil for future research.

\item {\bf Emission}. We could also study the emission case (we have been studying the absorption case); where $f_L$ and $f_R$ are made up of right moving and left moving Minkowski modes respectively.  These would posit a common observer in the future. The moving mirror \cite{fulling1976radiation} exemplifies aspects of both the ``Emission'' and ``Single Particle Model'' cases. These cases involve additional complications beyond the scope of our short note and are a topic of future research.
\end{itemize}

For the remainder of this note, we focus on the function $f_R$ in the right wedge without explicit reference to $f_L$, operating under the assumption that the localized driving source in the right wedge arises from an entangled paired source in Minkowski space.

\subsection{Sub-Wedge Causal Chain} \label{sec:chain_sources}

The sub-wedge sequence mentioned in equation (\ref{chain}) has an interesting interpretation.  The vacua are entangled from each wedge to the next sub-wedge, and each sub-wedge has vacuum particle expectations of the same character. In our construction, this is understood as a causal chain from Paz to Wright.  Each sub-wedge inherits a source $J$ and $f_R$ from the previous. This is also related to \cite{Svidzinsky2024MinkowskiVEA} where we also see causal chains from a common source.

\section{Localization} \label{sec:loc}

Up to now, our modes have been perfectly sharp in Rindler frequency. We now consider various within wedge localization techniques. From this point we use ``Rindler'' and ``Unruh'' modes interchangeably when restricted to a single wedge, as our focus is the localized support.

\subsection{Localization via Translated Wedge Inclusion}

\begin{figure}[h]
  \centering
\includegraphics[scale=0.4]{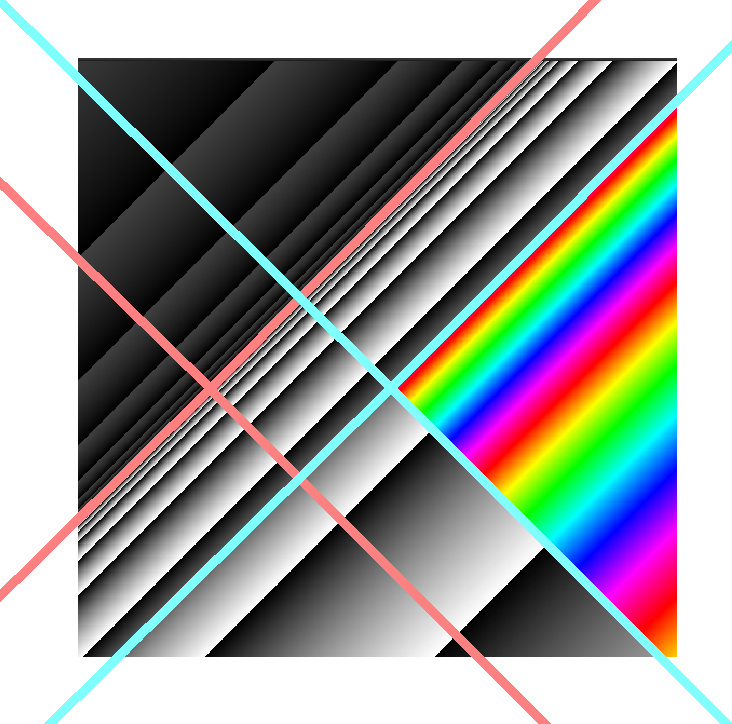}
\caption{A Wedge $W_c$ (cyan) inside of the wedge $W_0$ (pink). Rindler mode $r_q$ of $W_0$ (gray-scale) restricted to $W_c$ (rainbow).}
\label{restrict}
\end{figure}

Consider the two nested Rindler wedges $W_c \subsetneq W_0$ shown in Figure \ref{restrict}. Let $r_q$ denote a Rindler mode associated with $W_0$, analytically continued to the entire Minkowski space. The gray-scale region indicates the full support of $r_q$, while the rainbow-colored segment shows its restriction to the sub-wedge $W_c$.

By considering the restriction of $r_q$ to $W_c$, we have partially localized the observer and the mode. The restriction cuts off the high-frequency content of $r_q$ near the future horizon\footnote{Similarly, $r_{-q}$ experiences suppression near the past horizon.} of $W_0$. The resulting mode still spans the full spatial extent of $W_c$, but avoids the highly oscillatory behavior near the horizons of $W_0$.  The localization is not complete, however, the observer can still be anywhere within the wedge $W_c$, and the corresponding modes $r_k$ still exhibit thermal characteristics because of low-frequency oscillations extending as $x \rightarrow \infty$.

\begin{figure}[h]
  \centering
\includegraphics[scale=0.6]{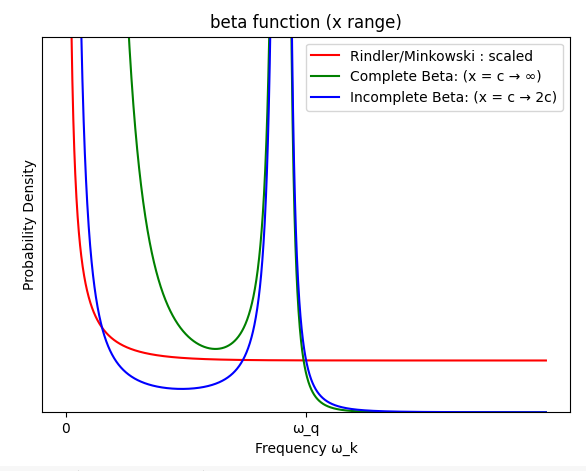}
\caption{The Rindler modes $r_k$ of $W_c$ show a peaked spectral overlap with $r_q$ at $\omega_k = \omega_q$ (green). The incomplete beta function (blue) strengthens the spectral peak and suppresses the peak at zero. Compare with purely thermal Minkowski/Rindler (red) curve.}
\label{peaked}
\end{figure}

To further study the situation we first review the Minkowski/Rindler situation by fixing $q$ and considering the modulus squared inner product $\left|\left<\varphi_q, r_k^{(c)} \right>\right|^2$ also known as the Bogoliubov $\left|\alpha^{(0 \rightarrow M)}_{kq}\right|^2$, from equation (\ref{bogoCM}). Using $|\Gamma(ib)|^2 = \frac{\pi}{b \sinh \pi b}$ we obtain the thermal response coming from the $\alpha_k$ term in
\begin{equation}
  \left|\left<\varphi_q, r_k^{(c)} \right>\right|^2 = \frac{1}{2\pi\omega_q} \left(\frac{1}{1 - e^{-\frac{2 \pi \omega_k}{a}}}\right)
\end{equation}
We graph a scaled version of this response as the red curve in Figure \ref{peaked}.

In contrast we next consider the modulus squared inner product $\left|\left<r_q^{(0)}, r_k^{(c)} \right>\right|^2$, also known as the Bogoliubov $\left|\alpha^{(c \rightarrow 0)}_{kq}\right|^2$, from equation (\ref{bogoC0}). We obtain
\begin{equation}
  \left|\left<r_q^{(0)}, r_k^{(c)} \right>\right|^2 = \frac{\sinh \frac{\pi \omega_q}{a}}{4\pi a (\omega_q - \omega_k) \sinh \pi \frac{\omega_q - \omega_k}{a} \sinh \frac{\pi \omega_k}{a}}
\end{equation}
as a function of $\omega_k$. The function exhibits a second-order pole at $\omega_k = \omega_q$, resulting in a sharply peaked feature, see the ``complete Beta'' green curve in Figure \ref{peaked}. Although the $\sinh$ terms encode aspects of the familiar thermal distribution, especially broadening near $\omega_k = 0$, the existence of the peak itself at $\omega_k = \omega_q$ originates from the geometric restriction, absent in the Minkowski/Rindler overlap.

\subsection{Diamond Localization via Reflected Wedge Intersection}

\begin{figure}[h]
  \centering
\includegraphics[scale=0.4]{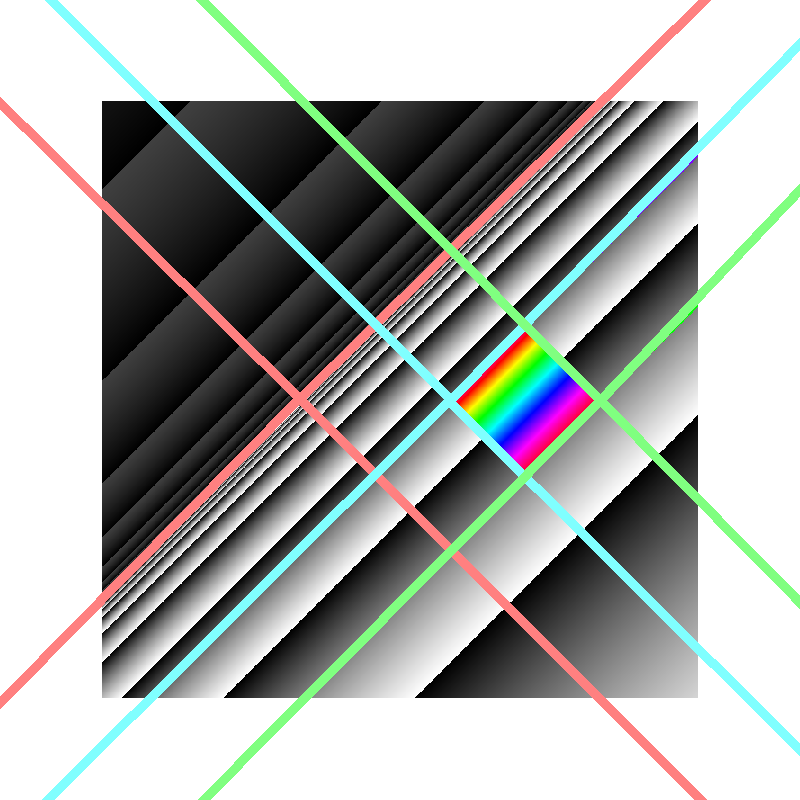}
\caption{The same situation as in Figure \ref{restrict} but we further intersect with a reflected (left) wedge $\widetilde{W}_{2c}$ (green). Rindler mode $r_q$ of $W_0$ (gray-scale) restricted to $W_c \cap \widetilde{W}_{2c}$ (rainbow).}
\label{diamond}
\end{figure}

Further localization results from intersecting $W_c$ with a reflected wedge $\widetilde{W}_{2c}$. This defines a more tightly localized diamond-shaped region, shown in Figure \ref{diamond}. The mode $r_q$ is now restricted to the intersection $W_c \cap \widetilde{W}_{2c}$, which eliminates much of the infrared behavior previously associated with the unrestricted wedge.

The Klein-Gordon inner product at $t=0$ now takes the form of an incomplete version of the beta function from equation (\ref{bogoC0}), corresponding to an integral\footnote{We could also use the other form of the beta function in equation (\ref{bogoTC0}) to compute the same inner product.} evaluated from $c$ to $2c$ rather than extending to infinity. This inner product does not however correspond to a mode expansion of the field, since we restrict the support to the diamond. So the construction does not define a complete orthonormal set, and cannot be used to build a full basis of field modes on the entire wedge.

Accordingly we instead interpret $r_q$ as part of a global mode expansion, we regard it as a compactly supported, non-invariant test function, i.e., a driving source localized to the diamond region, consistent with the source framework introduced in Section \ref{sec:drive}. We turn on $r_q$ exactly for a fixed period of $x-t$ (or $x+t$ for $r_{-q}$). The resulting spectral response in the diamond, computed from this truncated integral, is shown\footnote{The complete (green) beta function curve is actually independent of the choice of translation $c$, it is the same curve for any sub-wedge space-like inclusion (see modular automorphism Section \ref{sec:mod}). In contrast, the incomplete beta function curve (blue) does depend on the endpoint ($2c$ is shown).} as the blue curve in Figure \ref{peaked}. The plot reveals that the main spectral peak at $\omega_k = \omega_q$ persists, while the thermal contribution near $\omega = 0$ is significantly attenuated.

\subsection{Thermal to Localized Interpolation}
To further probe how global thermal structure transitions into a localized excitation, we consider the behavior of Rindler-to-Minkowski Bogoliubov coefficients when weighted by a Gaussian envelope. This allows us to interpolate between delocalized (thermal) and localized (spectrally peaked) behavior. We use parabolic cylinder functions \cite{AbramowitzStegun1964,Olver1959UniformAE} which are the analytic continuation of 

\begin{equation}
D_\nu(-z) = \frac{e^{-\frac{1}{4}z^2}}{\Gamma\left(-\nu\right)} \int_0^\infty  \dv{s} e^{zs} s^{-\nu - 1} e^{-\frac{1}{2} s^2}, \Re\nu < 0,
\end{equation}
where we use $-z$ instead of the usual $z$ so that future equations become simpler.

Without loss of generality, let $N_{\mu, \sigma} = e^{-\frac{1}{2} \frac{(x-t-\mu)^2}{\sigma^2}}$ be a (left-moving) Gaussian kernel with fixed $\mu$. We will multiply $\varphi_q^*$ by $N_{\mu, \sigma}$, but we could just as easily multiply $r_k$ by $N_{\mu, \sigma}$ for the same effect. Here the resulting Minkowski modes are treated as driving sources rather than elements of an orthonormal mode expansion. Since we are not working within an orthonormal mode expansion, the normalization of $N_{\mu, \sigma}$ is left implicit. This setup may be non-standard so we provide the explicit calculations.  We will examine the $N_{\mu,\sigma}$ modification of $\beta^{(c \rightarrow M)}_{kq}$ in  equation (\ref{bogoCM})

\begin{equation}
  \begin{aligned}
    \left< \varphi_q^* N_{\mu,\sigma}, r_k\right> &= \frac{1}{4\pi \sqrt{\omega_q \omega_k}} 2i \int_{\Sigma_W} e^{-i(\omega_q t - q x)} e^{-\frac{1}{2} \frac{(x-t-\mu)^2}{\sigma^2}} \partial_t (a(x-t))^\frac{i\omega_k}{a} \\
    &= \frac{1}{2\pi} \sqrt{\frac{\omega_k}{\omega_q}} a^{\frac{i \omega_k}{a} - 1} \int_0^\infty  \dv{x} e^{-\frac{1}{2} \frac{(x-\mu)^2}{\sigma^2} + i q x} x^{\frac{i\omega_k}{a} - 1} \\
    &= \frac{1}{2\pi} \sqrt{\frac{\omega_k}{\omega_q}} a^{\frac{i \omega_k}{a} - 1} \int_0^\infty \dv{x} e^{\left(-\frac{1}{2\sigma^2}\right) x^2 + \left(\frac{\mu}{\sigma^2} + i q \right) x + \left( -\frac{\mu^2}{2\sigma^2}\right)} x^{\frac{i\omega_k}{a} - 1}  \\
    &= \frac{1}{2\pi a} \sqrt{\frac{\omega_k}{\omega_q}} e^{-\frac{\mu^2}{2 \sigma^2}} \sigma^{\frac{i\omega_k}{a}} a^{\frac{i \omega_k}{a} }  \int_0^\infty \dv{s} e^{(\frac{\mu}{\sigma} + i q \sigma)s} s^{\frac{i\omega_k}{a} - 1} e^{-\frac{1}{2} s^2} \\
    &= \frac{1}{2\pi a} \sqrt{\frac{\omega_k}{\omega_q}} e^{-\frac{\mu^2}{2 \sigma^2}} {(\sigma a)}^{\frac{i \omega_k}{a}} e^{\frac{1}{4}(i q \sigma + \frac{\mu}{\sigma})^2} \Gamma\left(\frac{i\omega_k}{a}\right) D_{-\frac{i\omega_k}{a}}(-i q\sigma - \frac{\mu}{\sigma}) \\
    &=  \frac{1}{2\pi a } \sqrt{\frac{\omega_k}{\omega_q}} e^{-\frac{\mu^2}{2 \sigma^2}}  (\sigma a)^\frac{i\omega_k}{a} e^{\frac{1}{4} z^2} \Gamma(-\nu) D_\nu(-z)
  \end{aligned}
\end{equation}
where $\Sigma_W$ is the Cauchy surface $\eta=0$ on the Rindler wedge $W$,  $x = \sigma s$, $z = i q \sigma + \frac{\mu}{\sigma}$, and $\nu = -\frac{i \omega_k}{a}$. And then
\begin{equation}
  \begin{aligned}
    \left|\left< \varphi^*_q N_{\mu,\sigma}, r_k \right>\right|^2 &= \frac{1}{2\pi a \omega_q} \frac{\omega_k}{2\pi a} e^{-\frac{\mu^2}{\sigma^2}} \left| e^{\frac{1}{2} z^2} \right| \left| \Gamma(-\nu) \right|^2 \left| D_\nu(-z) \right|^2 \\
  \end{aligned}
\label{pcf}
\end{equation}
\begin{figure}[h]
\centering
\includegraphics[scale=0.5]{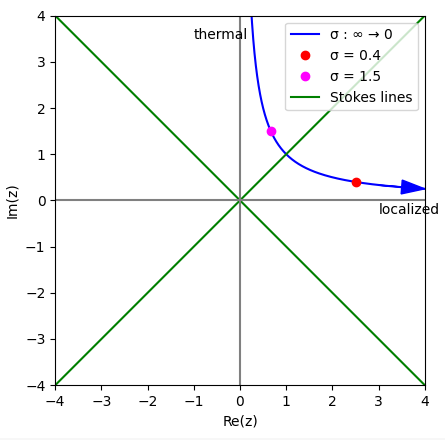}
\caption{Trajectory of $z = i q \sigma + \frac{\mu}{\sigma}$ as $\sigma$ interpolates between thermal and localized regimes. For $\sigma$ starting at $\infty$, the trajectory starts at the positive infinite imaginary axis, aligning with the dominant thermal component of the excitation. As $\sigma \to 0$, the system crosses a Stokes line and transitions into a sharply localized, source-driven configuration, where thermal character disappears. See also corresponding Figure \ref{pcf_sigma_curves}.}
\label{stokes}
\end{figure}
\begin{figure}
\centering
\includegraphics[scale=0.5]{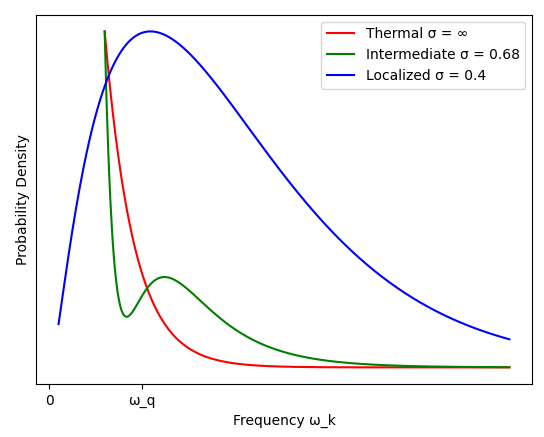}
\caption{$\left|\left< \varphi^*_q N, r_k \right>\right|^2$ for various values of $\sigma$ (probability density is scaled for comparison). $\omega_q = 1$, $q = 1$, $a=1$, $\mu = 1$. See also corresponding Figure \ref{stokes}.}
\label{pcf_sigma_curves}
\end{figure}
From \cite{Olver1959UniformAE} we have
\begin{equation}
  D_\nu(-z) = e^{-i\pi\nu}z^{\nu}e^{-\frac{1}{4}z^2} \{ 1 + O(|z|^{-2}) \} + \frac{(2 \pi)^{\frac{1}{2}}}{\Gamma(-\nu)} z^{-\nu - 1} e^{\frac{1}{4}z^2} \{1 + O(|z|^{-2})\}
\label{asym}
\end{equation}
when $-\frac{1}{4}\pi + \epsilon \le \arg z \le \frac{3}{4} \pi - \epsilon$.

The two asymptotic regimes correspond to physically distinct interpretations: The second $e^{\frac{1}{4}z^2}$ term dominates for $z \to \infty$ as $\sigma \to 0$ and the first  $e^{-\frac{1}{4}z^2}$ term dominates for $z \to i \infty$ as $\sigma \to \infty$.  This is a Stokes phenomenon\footnote{See \cite{hashiba2021stokes} for a similar situation where the Stokes phenomenon is applied to particle production in simple expanding backgrounds, preheating after $R^2$ inflation, and a transition model with smoothly changing mass.} which flips over as we cross the Stokes line at $\arg z = \frac{\pi}{4}$. The situation is pictured in Figure \ref{stokes}.

We next combine equations (\ref{pcf}) and (\ref{asym}).  First for the thermal part that comes from the $e^{-\frac{1}{4}z^2}$ term where $\sigma \to \infty$ we have
\begin{equation}
  \begin{aligned}
    2\pi a \omega_q \left|\left< \varphi^*_q N, r_k \right>\right|^2 &= \frac{\omega_k}{2 \pi a} e^{-\frac{\mu^2}{\sigma^2}} \left|e^{-i\pi \nu} z^\nu\Gamma\left(\frac{i\omega_k}{a}\right)\right|^2   \\
    &= \frac{\omega_k}{2 \pi a} e^{-\frac{\mu^2}{\sigma^2}}  e^{\frac{-2\pi \omega_k}{a}} \left|e^{\frac{-2i\omega_k}{a} \log{(\frac{\mu}{\sigma} + iq\sigma)}}\right|^2  \frac{\pi}{\frac{\omega_k}{a} \sinh \frac{\pi \omega_k}{a}} \\
  &\to  e^{\frac{-2\pi \omega_k}{a}} \left|e^{\frac{-2i\omega_k}{a} \log{i}}\right|^2  \frac{1}{2\sinh \frac{\pi \omega_k}{a}} \\
  & =  e^{\frac{-2\pi \omega_k}{a}} e^{\frac{\pi \omega_k}{a}}  \frac{1}{ \left( e^{\frac{\pi \omega_k}{a}} - e^{\frac{-\pi \omega_k}{a}} \right)} \\
  & =  \frac{1}{e^{\frac{2\pi\omega_k}{a}} - 1} \\
  \end{aligned}
\end{equation}
which we expect by construction. For the localized part that comes from the $e^{\frac{1}{4}z^2}$ term where $\sigma \to 0$ we have
\begin{equation}
  \begin{aligned}
    2\pi a \omega_q \left|\left< \varphi^*_q N, r_k \right>\right|^2 &= \frac{\omega_k}{a} e^{-\frac{\mu^2}{\sigma^2}} \left|e^{z^2}z ^ {2(-\nu - 1)}\right| \\
    &= \frac{\omega_k}{a} e^{-\frac{\mu^2}{\sigma^2}} \left|e^{ \left(  iq\sigma + \frac{\mu}{\sigma} \right)^2 } e^{{2\left(\frac{i\omega_k}{a} - 1\right)\log{ \left(  iq\sigma + \frac{\mu}{\sigma} \right)   }   }   }\right| \\
    &\to \frac{\omega_k}{a} e^{\frac{-2\epsilon\omega_k}{a}} f(\sigma,\mu) \\
  \end{aligned}
\end{equation}
where the thermal pole at zero has disappeared. While the precise asymptotic form is not critical, we can control the ultraviolet behavior by taking $z$ to $(1+i\epsilon)\infty$ which remains within the localized Stokes region. This introduces a regulating factor of the form $e^{-2\epsilon \omega_k / a}$, which suppresses high-frequency contributions.

We restrict attention to small but finite $\sigma$, and $\sigma \rightarrow 0$ corresponds to a vanishing source rather than a delta function. We focus on the small-but-nonzero $\sigma$ regime where the thermal character is already suppressed. See Figure \ref{pcf_sigma_curves} for representative plots across varying values of $\sigma$.

\section{Future Research} \label{sec:future}

Several directions naturally extend the present work:
\begin{itemize}
\item {\bf Moving Mirrors}. A moving mirror can be viewed as both a detector and a source (Section \ref{sec:drive}); in this work the source and detector roles were treated separately. Extending the framework to include moving mirror localizations is an open problem.
\item {\bf Detector Modeling}. In the present framework, excitations are injected directly at the source. These source-driven excitations are expected to constrain detector-based models, but a full treatment lies beyond the scope of this paper.
\item {\bf Mode Pairing}. The construction of $f_L$ in Section \ref{sec:drive} was not pursued in Section \ref{sec:loc}. A systematic analysis of the $f_L$–$f_R$ pairing could clarify correlation structures more generally.
\item {\bf Alternative Driving Source Models}. Beyond entangled emission, models with single-particle sources or time-reversed scenarios (Section \ref{sec:drive_phy}) may provide finer control over entanglement structure.
\item {\bf General Fields}. Extending the analysis to massive fields or to full QED would enable the study of localized virtual particle–antiparticle pair production as well as optical photons, complementing the massless case.


\end{itemize}

%

\section{Conclusion} \label{sec:conc}

We have presented a framework in which individual microstates of the thermal Unruh ensemble can be realized as physically localized source-driven excitations, making the role of acceleration explicit through specific constructions rather than purely kinematic. Our construction complements the standard detector-based interpretation by showing how some apparent thermality can coexist with the microstate-level descriptions. We analyzed the wedge-restricted Rindler/Unruh modes, which reveal a mixture of thermal and non-thermal features.  Finally, we studied compact wave-packets via parabolic cylinder functions, providing a smooth interpolation between global thermal Rindler modes and localized non-thermal excitations.

The framework suggests several directions for further investigation, including systematic mode pairing, moving mirror analogues, generalizations to massive fields, and potential extensions to curved spacetimes, with applications to near-horizon excitations in black hole physics.

\section{Acknowledgments}
I thank Frodden and Valdés for their excellent exposition \cite{frodden2018unruh}, and Beisert for his insightful lecture notes \cite{beisert2012quantum}. I'm also grateful to Ben Commeau, Daniel Justice, Edward Randtke, and ChatGPT for helpful discussions.

\bibliographystyle{ieeetr}
\bibliography{bibliography}

\end{document}